\documentstyle[times,rotating]{jaa}
\begin{document}
\title[North-South Distribution of Solar Flares during Cycle 23]
{North-South Distribution of Solar Flares during Cycle 23}
               
\author[Bhuwan Joshi et al]
       {Bhuwan Joshi\thanks{e-mail:bhuwan@aries.ernet.in} $^1$, 
        P. Pant $^1$ and P. K. Manoharan$^2$\\ 
        $^1$Aryabhatta Research Institute of Observational Sciences, 
        Nainital 263 129, India\\
        $^2$Radio Astronomy Centre, TIFR, Ooty 643 001, India}
\maketitle
\label{firstpage}
\begin{abstract}
In this paper, we investigate the spatial distribution of solar flares 
in the northern and southern hemisphere of the Sun that occurred during 
the period 1996 to 2003.
This period of investigation includes the ascending phase, 
the maximum and part of descending phase of solar cycle 23. 
It is revealed that the flare
activity during this cycle is low compared to previous solar cycle,
indicating the violation of Gnevyshev-Ohl rule. 
The distribution of flares with respect to heliographic latitudes
shows a significant asymmetry between northern and southern hemisphere 
which is maximum during the minimum phase of the solar cycle.
The present study indicates that the activity dominates the northern hemisphere 
in general during the rising phase of the cycle (1997-2000).  
The dominance of northern
hemisphere is shifted towards the southern hemisphere after the
solar maximum in 2000 and remained there in the successive years.
Although the annual variations in the asymmetry time series during cycle 23 
are quite different from cycle 22, they are comparable to cycle 21. 


\end{abstract}

\section{Introduction}
\label{sec:intro}
The distribution of various solar activity features with respect to
heliographic latitudes as a function of time has been investigated in  
several studies. These activity features include flares, filaments, magnetic
flux, sunspot numbers, sunspot area etc. These studies indicate that a 
solar cycle is not symmetric considering the distribution of solar activity  
separately in northern and southern hemisphere. This intrinsic feature
(N-S asymmetry) poses a challenge for dynamo model calculations.  

Howard (1974) examined the N$-$S distribution of solar magnetic flux for the 
period 1967$-$1973 and concluded that about 95 \% of the total magnetic flux 
of the Sun is confined to latitudes below 40$^{o}$ in both the hemispheres.  
It was also found that total magnetic flux in the north exceeded to that 
in the south by 7 \%. Roy (1977) studied the N$-$S distribution in the data of major flares, sunspot area 
and their magnetic configuration during 1955-1974 and 
found the dominance of northern hemisphere over 
the southern one in all these categories. 
Garcia (1990) investigated the N$-$S distribution of soft X-ray flares (class $\ge$ M1) during solar cycle
20 and 21. It was concluded that the spatial distribution of flares varies within a solar cycle such 
that the preponderance of flares occurs in the north during the early part of the cycle and then moves 
south as the cycle progresses. 
Kno\v{s}ka (1985) investigated the N-S asymmetry of H$\alpha$ flare index, introduced by Kleczek (1952), for 
cycles 17-20, which was later extended by Ata\c{c} and \"{O}zg\"{u}\c{c} (2001) till cycle 22. Their results show
a long-term periodic behavior in asymmetry time series.
Verma (1993) examined the N-S asymmetry of various solar active phenomena
and reported cyclic behavior of asymmetry. 
Joshi (1995) and Li et al. (1998) studied the N$-$S asymmetry 
of H$\alpha$ and soft X-ray flares respectively during solar cycle 22. In both the above investigations
a southern dominance was prevalent. Joshi and Pant (2005) analysed the data of solar H$\alpha$ flares
during solar cycle 23 to investigate their spatial distribution.
Recently Knaack et al. (2004, 2005) studied the temporal and spatial 
variations in the photospheric magnetic flux between the northern and southern hemisphere of the 
Sun from 1975 to 2003 and reported significant periodic variations of magnetic activity between the two 
hemispheres. They also studied N-S asymmetry using the monthly averaged sunspot areas obtained
for the period of 1874-2003.

In the present study, we investigate the latitudinal distribution 
of H$\alpha$ flares during the present solar cycle 23. We have studied the 
yearly variations in N-S asymmetry and discussed the significance of observed asymmetry
by using the binomial probability test. These results for solar cycle 23 have been
compared with the behavior of previous cycles 22 and 21 as reported by Temmer et al. (2001).

\begin{figure}
\begin{center}
\includegraphics[width=8.7cm, height=8cm]{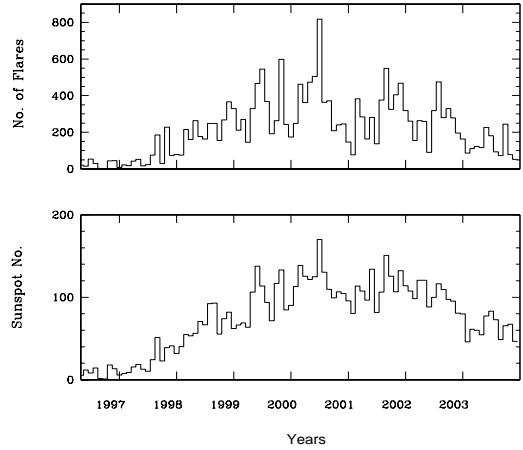}
\vspace{-1.2cm}
\caption{Monthly numbers of H$\alpha$ flares and monthly mean sunspot numbers from 1996-2003.}
\end{center}
\end{figure}

\section{H$\alpha$ flare data}
The data used in the 
present study have been collected from H$\alpha$ flare lists published in 
the SGD (Solar Geophysical Data) during the time span of 01 May 1996 to 
31 December 2003, covering almost 8 years of solar cycle 23. During 
this period the occurrence of 20235 H$\alpha$ flares {\bf is} reported. 
In H$\alpha$, flares are classified according to their importance and 
brightness classes. The importance class (S=subflare, or 1, 2, 3, or 4 for 
successively large flares) denotes the size of the flare and the brightness 
class (f=faint, n=normal, b=bright) corresponds to a subjective estimate 
of the intensity of the emission. In the list of H$\alpha$ flares, 
classifications  are not given for some events. After excluding such events,
we get a total of 20223 H$\alpha$ flares, which provides an extensive database 
for the present 
study. Figure 1 shows the plot of monthly flare counts as well as monthly mean 
sunspot numbers during the period of our investigation.

\section{Latitudinal distribution and N-S Asymmetry}
\begin{table}
\caption{Number of H$\alpha$ flares at different latitude bands in the
northern (N) and southern (S) hemispheres are tabulated for each year.
The binomial probability 
(Prob.) and the dominant hemisphere (DH) is given for all the years as well as
for all the latitudinal bands. Dash ($-$) represents that the probability is 
not significant. Flares occurred exactly
at the equator have been excluded.}
\begin{center}
\rotatebox{270}{
\begin{tabular}{cccccccc|c|c}\hline
Years& \multicolumn{6} {c} {Number of flares}& &Prob.&DH\\ \cline{2-8}
   &0-10$^{o}$&10-20$^{o}$&20-30$^{o}$&30-40$^{o}$&40-50$^{o}$&\(>\)50$^{o}$&Total&&\\
\hline
1996 N   &21   &2   &0   &1  &0  &0  &24  &1.107$\times$10$^{-33}$ &S  \\
~~~~~~ S &112  &53  &21  &2  &0  &0  &188 &                        &   \\
1997 N   &39   &160 &230 &20 &0  &0  &449 &5.641$\times$10$^{-6}$  &N  \\
~~~~~~ S &5    &103 &199 &18 &1  &1  &327 &                        &   \\
1998 N   &13   &635 &503 &54 &1  &1  &1207&0.428                   &$-$\\
~~~~~~ S &3    &366 &807 &31 &9  &0  &1216&                        &   \\
1999 N   &169  &1240&811 &105&11 &0  &2336&7.017$\times$10$^{-30}$ &N  \\
~~~~~~ S &71   &874 &644 &38 &0  &0  &1627&                        &   \\
2000 N   &463  &1327&633 &55 &1  &2  &2481&8.712$\times$10$^{-14}$ &N  \\
~~~~~~ S &248  &1288&373 &79 &1  &0  &1989&& \\
2001 N   &503  &978 &262 &5  &1  &0  &1749&0.058                   &S  \\
~~~~~~ S &449  &1062&282 &42 &8  &0  &1843&                        &   \\
2002 N   &250  &746 &255 &6  &0  &0  &1257&1.329$\times$10$^{-35}$ &S  \\
~~~~~~ S &675  &1006&272 &4  &0  &0  &1957&                        &   \\
2003 N   &407  &315 &14  &3  &0  &0  &739 &0.069                   &S  \\
~~~~~~ S &222  &475 &83  &17 &0  &0  &797 &                        &   \\
\hline
Total N  &1865&5403&2708 &249&14  &3  &10242&0.018                 &N  \\
~~~~~~ S &1785&5227&2681 &231&19  &1  &9944 &  &\\
\hline
Prob.    &0.093&0.956 &0.356&0.205&0.189&0.125&0.018&&    \\
         &     &      &     &    &    &    && &\\
\cline{1-8}
DH      &N     &$-$   &$-$  &$-$ &$-$ &$-$ &N&&    \\
\hline
\end{tabular}
}
\end{center}
\end{table}
\label{sec:using}
To study the spatial distribution of flares with respect to heliographic
latitudes, we have calculated the number of flares in the interval of 10$^{o}$
latitude for northern and southern hemispheres (see Table 1). 
In this Table those events have been excluded which occurred
at 0$^{o}$ latitude. 
Since the number flares above 50$^{o}$ latitude are very small
in both the hemispheres, the number of flares occurring above 50$^{o}$ latitude is merged in one 
group. 
Column 8 of Table 1 gives the total number of flares in the northern and
southern hemispheres. It is evident from Table 1 that for all these years one 
hemisphere produces more flares than the other.

It is customary to describe N-S asymmetry by an asymmetry index

\begin{equation}
A= \frac{{N} - {S}}{{N} + {S}},
\end{equation}

where N and S are the yearly number of flares in northern and southern 
hemisphere of the Sun respectively. 
The asymmetry indices, based on annual flare counts from 1996 to 2003,
have been plotted in Figure 2. The statistical significance of the flare 
dominance in northern and southern hemispheres has been assessed by using the
binomial probability distribution. Let us consider a distribution of $n$ objects in 2 classes. The binomial formula
gives us the probability $P(k)$ of getting $k$ objects in class 1 and $(n-k)$ objects
in class 2, such that
\begin{equation}
P(k) =  \frac{n!}{k!(n-k)!} p^{k} (1-p)^{n-k}
\end{equation}
and the probability to get more than $d$ objects in class 1 is given by
\begin{equation}
P(\ge d) =  \sum_{k=d}^{n} P(k). 
\end{equation}
In general, when P($\ge$d) $>$ 10 $\%$, implies a statistically insignificant
result (flare activity should be regarded as being equivalent for the two hemispheres) , 
when 5$\%$ $<$ P($\ge$d) $<$ 10$\%$ it is marginally significant, and when
P($\ge$d) $<$ 5$\%$ we have a statistically significant result (flare occurrence is not due to random 
fluctuations). The calculated values of probability is given in Table 1 and
based on that the highly significant, marginally significant and insignificant 
values of asymmetry indices are marked with different symbols in Figure 2.
Figure 2 also shows the yearly N-S asymmetry (given by equation 1) 
of H$\alpha$ flare index (dashed line) from 1996 to 2003.
\begin{figure}
\begin{center}
\includegraphics[height=7cm, width=10cm]{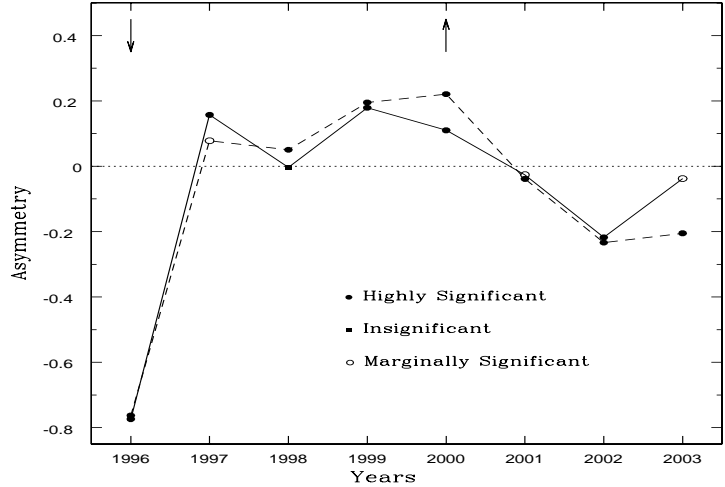}
\caption{Plot of annual N-S asymmetry index of H$\alpha$ flare counts 
(solid line) and H$\alpha$ flare index (dashed line) from 
1996 to 2003. Arrows indicate the maximum and minimum phases, respectively, of solar cycle 23.}
\end{center}
\end{figure}

\section{Discussions and conclusions}
\label{sec:frontmatter}
Temmer et al. (2001) and Temmer (2004) made an extensive statistical analysis of H$\alpha$ flare 
data of cycle 21 and 22 to study their temporal behavior and spatial distribution over the solar cycle.
The comparison of monthly flare counts during cycle 23 (figure 1) with those reported by Temmer et al, (2001) for
cycle 22 and 21 shows that the activity level during cycle 23 is significantly lower than two previous cycles. 
This clearly indicates 
the violation of the Gnevyshev-Ohl (G-O) rule in terms of level of flare activity, for the pair of solar cycles 22-23. This empirical rule states that the sum of sunspot numbers over an odd cycle exceeds that of 
the preceding even cycle (Gnevyshev and Ohl 1948). The sunspot numbers, starting from cycle 0 (i.e., from the year 
1750), show that G-O breakdown had also occurred for the Hale cycles consisting of the 11 year pair of cycles 4-5
and 8-9. Komitov and Bonev (2001) examined the conditions for the violation of the G-O rule. They
analysed a long data set of 152 solar activity cycles obtained from direct and indirect records and pridicted 
a high probability for violation of the G-O rule for the pair of the cycles 22-23.

Table 1 shows several interesting aspects of flare distribution with the evolution of solar cycle. In the beginning
of the cycle 0-10$^{o}$ latitudinal belt in southern hemisphere produced maximum number of flares, which 
could be a remnant of the preceding cycle. In the year 1997, just after solar minimum, most of the flares were produced
in the 20-30$^{o}$ latitudinal belt and with the progress of solar cycle, the flare occurrence increased
in lower latitudes also. Table also shows that 10-20$^{o}$ latitudinal belt was the highest 
flare producing region. 

Figure 2 shows the annual variations in N-S asymmetry. There is a strong southern dominance during the
solar cycle minima in 1996. This behavior of asymmetry that it peaks at or around the minimum phase of solar activity
has been reported in several studies with different manifestations of solar activity (Swinson, Koyama, and Saito 1986;
Vizoso and Ballester 1990; Joshi and Joshi 2004). 
In 1997, 1999 and 2000, when the cycle was in ascending phase, northern hemisphere dominated. 
The preference for northern hemisphere during the rising and maximum phase of cycle 23 is reported
by Ata\c{c} and \"{O}zg\"{u}\c{c} (2001) and Joshi and Joshi (2004) in yearly values of H$\alpha$ and soft X-ray 
flare index respectively. Gopalswamy et al. (2003) compared the latitudinal distribution of prominence 
eruptions (PEs) and coronal mass ejections (CMEs) as a function of time from 1996 to 2002. Their study
revealed that there is a shift in the dominance of PEs and CMEs activity from northern to southern
hemisphere after solar maximum in 2000. Similar trend has been found in the present investigation 
with solar flare count data.
The yearly variations in N-S asymmetry for flare counts and flare 
index (Figure 2) show a similar trend. In the year 2000 flare index asymmetry 
is stronger in northern hemisphere compared to flare count asymmetry, 
which indicates that most of the major flares occurred in northern part of the 
solar disk during solar cycle maxima. Similarly during October and November
months of 2003, most of the big flares occurred in southern hemisphere of Sun
and consequently  the asymmetry in flare index values became 
stronger in southern part of solar disk.
Comparing our results with Temmer et al. (2001) we find that the variations in N-S asymmetry index
during cycle 23 differ from cycle 22 but are similar to cycle 21. There was mostly 
a southern dominance during cycle 22 while cycle 21 showed northern dominance during the early phases and 
southern dominance during the later phases. The present investigation is consistent with the N-S asymmetry
analysis performed with soft X-ray flare index data during cycle 21-23 (Joshi and Joshi 2004).

\section{Acknowledgements}
We acknowledge the constructive comments and suggestions from an anonymous referee which improved the scientific contents of the paper. Flare Index data used in this study were calculated by T. Ata\c{c} and A. \"{O}zg\"{u}\c{c} from Bogazici University Kandilli Observatory, Istanbul, Turkey.

\end{document}